\documentclass[preprint,12pt]{elsarticle}
\usepackage{graphics}
\usepackage{rotating}
\usepackage{mathrsfs}
\usepackage{amsmath}
\usepackage{lscape}
\usepackage[usenames,dvipsnames]{color}
\usepackage{tikz,ifthen,calc}
\usetikzlibrary{snakes,shapes}
\usetikzlibrary{patterns}
\usetikzlibrary{plotmarks}
\usetikzlibrary{calc}
\usepackage{pgfplots}

\newcommand{\be}{\begin{equation}}
\newcommand{\ee}{\end{equation}}
\newcommand{\bea}{\begin{eqnarray}}
\newcommand{\eea}{\end{eqnarray}}

\journal{}
\begin{document}
\begin{frontmatter}

\title{Double binding energy differences: Mean-field or pairing effect?}
\author{Chong Qi\corref{email}}
\cortext[email]{Corresponding author.\\
\textit{E-mail address:} chongq@kth.se (Chong Qi)}
\address{Royal Institute of Technology (KTH), Alba Nova University Center,
SE-10691 Stockholm, Sweden}

\begin{abstract}
In this paper we present a systematic analysis on the average interaction between the last protons and 
neutrons in atomic nuclei, which can be extracted from the double differences 
of nuclear binding energies. The empirical average proton-neutron interaction $V_{pn}$ thus derived from experimental data can be described in a very simple form as the interplay of the nuclear mean field and the pairing interaction.
It is found that the smooth behavior as well as the local fluctuations of the $V_{pn}$ in even-even nuclei with $N\neq Z$ are dominated by the contribution from the proton-neutron monopole interactions.
A strong additional contribution from the isoscalar monopole interaction and isovector proton-neutron pairing interaction is seen in the $V_{pn}$ for even-even $N=Z$ nuclei and for the adjacent odd-$A$ nuclei with one neutron or proton being subtracted.
\end{abstract}


\end{frontmatter}

The binding energies of atomic nuclei reflect the interactions of its two constitutes, protons and neutrons. 
Differences of binding energies give nuclear separation
energies which can be used to isolate specific correlations. 
The zigzag behavior of one-body separation energies has long been well known. It
provides clues to the pairing correlation between like nucleons~\cite{bm,Ber09}.
In past decades, the structure of nuclei has been understood to a large extent 
within a mean-field (single-particle potential in the Hartree-Fock or particle-hole channel) plus pairing approach. 
Modern nuclear structure model calculations within this framework can reproduce the binding energies
of nuclei over the whole nuclear chart with a high precision~\cite{Gor09,Ben03,Lunney}.

The correlation between the proton and the neutron has been expected to play a key role in
the development of collective correlation~\cite{Fed77,Cas06} and in the evolution of
the shell structure~\cite{Otsuka01,Sor08}.
The (phenomenological) average interaction between the
last protons and the last neutrons in even-even
nuclei can be extracted from the double difference of binding
energies as~\cite{Zhang89}
\begin{eqnarray}\label{vpn-ee}
\nonumber V_{pn} {(Z, N)}&=& \frac{1}{4}\left[
B(Z,N)+B(Z-2,N-2)\right. \\ 
&&- \left.B(Z-2,N)-B(Z,N-2) \right],
\end{eqnarray}
where $B(Z,N)$ is the (positive) binding energy of a nucleus with $Z$ protons and $N$ neutrons. 
The factor $1/4$ takes into account the fact that four additional pairs are 
formed by the last two protons and neutrons. 

Recently, the proton-neutron interaction has attracted renewed interest, which may reveal additional
nuclear structure effects~\cite{Cas05,Cas06a,Cas06b,Sto07,Fu10,Jiang12}
and shed light on the possible existence of novel pairing correlation modes~\cite{Macc00,Good01,Satu01,Chas07,Qi11}.
Stoitsov {\it et al.} showed that the global 
properties of $V_{pn}$ can be reproduced by 
Hartree-Fock-Bogoliubov (HFB) calculations with the Skyrme functional plus a density-dependent $\delta$ pairing 
interaction~\cite{Sto07}. A detailed calculation was also done in Ref. \cite{Ben11} where the effects of the deformation and collective fluctuation on $V_{pn}$ were analyzed. It would be interesting to understand the microscopic mechanism behind the success of these calculations and to explore the extent to which the empirical $V_{pn}$ can be incorporated into nuclear models where only proton-proton and neutron-neutron pairing correlations
are explicitly taken into account. In this work I make an attempt in this direction by separating the contributions from the pairing and the mean field upon $V_{pn}$ in a simple way.
But perhaps even more appealing is 
to explore the local fluctuations of $V_{pn}$ around the average values which large-scale HFB calculations 
fail to explain~\cite{Sto07}. These fluctuations may carry further
nuclear structure information and serve as a constraint in future developments of nuclear structure models.

\begin{figure}
\centerline{\includegraphics[width=0.65\textwidth]{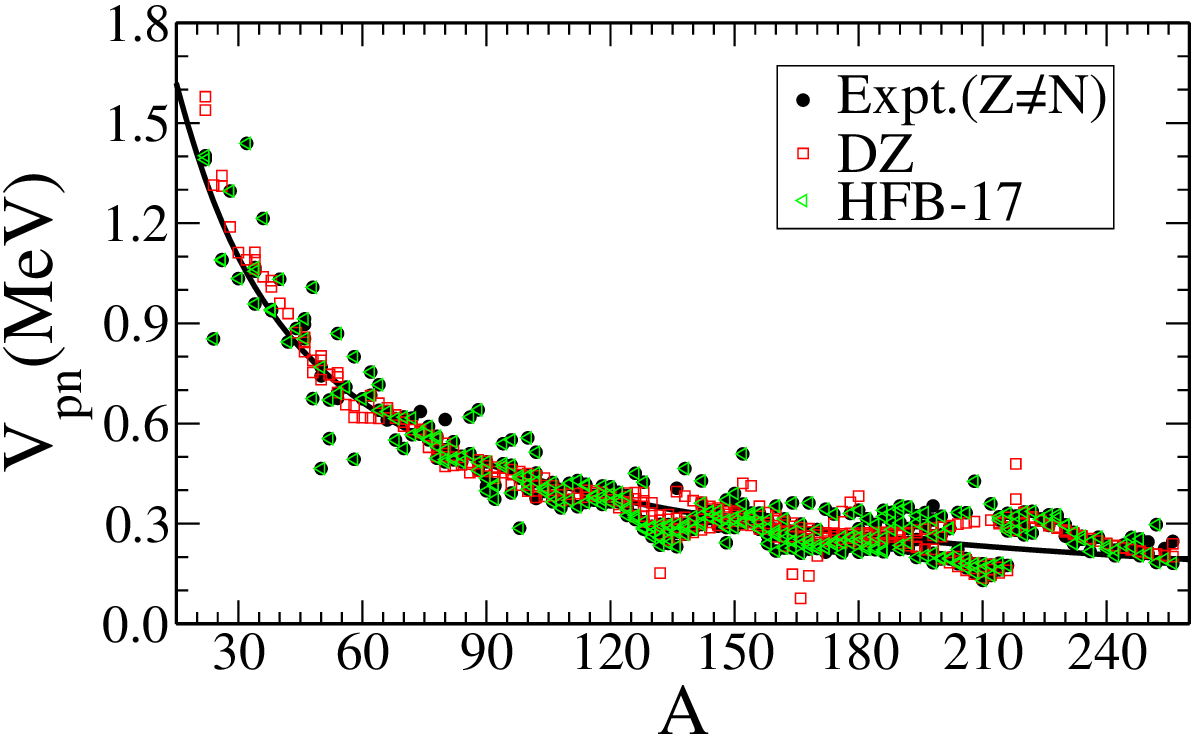}}
\caption{(color online). Empirical proton-neutron interactions in even-even nuclei
extracted from experimental nuclear masses \cite{Audi03} as a function of the mass number $A$.
Open symbols correspond to calculations with the DZ \cite{Duf95} and HFB-17 \cite{Gor09} mass models.\label{ee}}
\end{figure}

The $V_{pn}$
extracted from experimental nuclear binding energies \cite{Audi03} (taken as positive values) are plotted in Fig. \ref{ee}. It can be seen
that $V_{pn}$ evolve rather smoothly as a function of mass number $A$. 
In fact, this average behavior of $V_{pn}$ also probes the symmetry energy term (i.e., the isospin-dependence 
of the binding energy) in the macroscopic mass formula \cite{Satu97}. The overall trend of $V_{pn}$ can be 
well approximated by a smooth relation of $(a+a_sA^{-1/3})/A$
 (see the solid line in Fig.~\ref{ee} that fits experimental data) \cite{Sto07}.

As a comparison,  in Fig.~\ref{ee} we also plotted the $V_{pn}$ calculated from the HFB-17~\cite{Gor09} and Duflo-Zuker (DZ)~\cite{Duf95,dz}  mass models. We take the calculated mass table from Ref. \cite{Gor09} for simplicity instead of repeating the large-scale HFB calculations done in Refs. \cite{Gor09,Sto07,Ben11}. The HFB-17 model  is comprised of the conventional 
Skyrme functional, a contact pairing force and a number of empirical corrections including the Wigner energy. It can reproduce experimental data within a deviation of 581~keV \cite{Gor09}. 
There are several versions of the DZ model available. Only the simplified DZ model is used in the present work. It contains ten terms and can reproduce experimental nuclear 
masses within a deviation factor 
of around 550~keV \footnote{One may argue that the DZ model has more than ten parameters since it also contains several phenomenal scaling factors and isospin-dependent terms.}.
It is seen from Fig. \ref{ee} that both calculations 
can reproduce nicely experimental data, mostly within a deviation of $20\%$. 

We note that Eq.~(\ref{vpn-ee}) can be rewritten as
\begin{eqnarray}\label{s2n}
\nonumber 4V_{pn} {(Z, N)}&=&S_{2n}(Z,N)-S_{2n}(Z-2,N)\\
&=&S_{2p}(Z,N)-S_{2p}(Z,N-2),
\end{eqnarray}
where $S$ denotes the separation energy.
From this relation it is easily seen that $ V_{pn}$ also measures the
extra binding gained by the neutron (proton) pair when two additional protons (neutrons) are added.

\begin{figure}
\centerline{\includegraphics[width=0.65\textwidth]{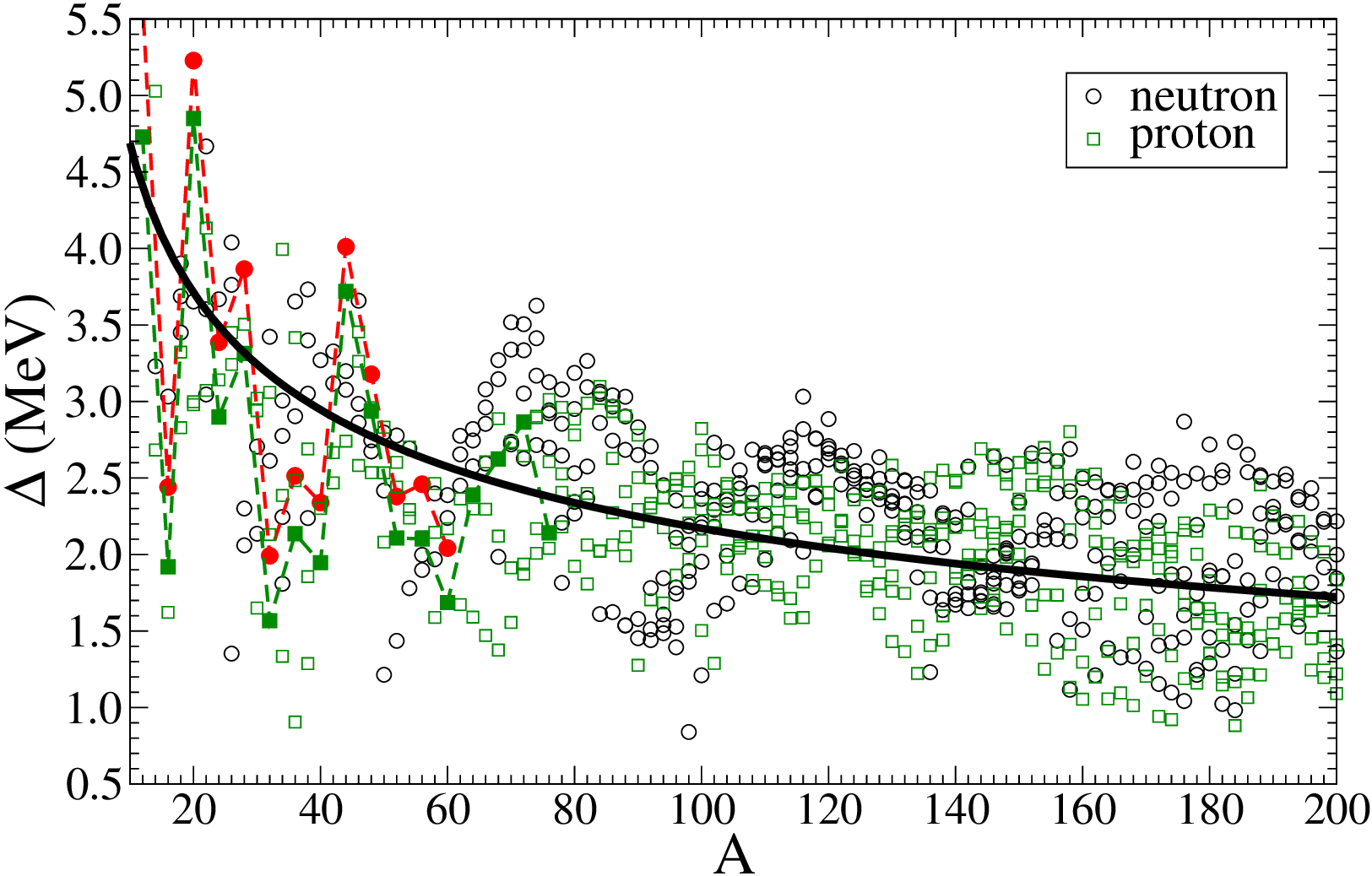}}
\caption{(color online). Empirical proton-proton (squares) and neutron-neutron (circles) interactions in even-even nuclei
extracted from experimental nuclear masses as a function of the mass number $A$ \cite{Audi03}.
The solid symbols denote those in the $N=Z$ nuclei.\label{pair}}
\end{figure}

The two-nucleon separation energies in even-even nuclei can be written as
\begin{eqnarray}
S_{2n}(Z,N)=2S_n(Z,N-1)+\Delta_n(Z,N),
\end{eqnarray}
where
\begin{equation}
\Delta_n(Z,N)=B(Z,N)+B(Z,N-2)-2B(Z,N-1).
\end{equation}
To a large extent $\Delta(Z,N)$ measures the pairing interaction between the last two 
like nucleons~\cite{Ber09,Satu98}. The proton and neutron $\Delta(Z,N)$ extracted from experimental binding energies are plotted in Fig. \ref{pair}. It is seen that the $\Delta(Z,N)$ values are mostly within $1-5$ MeV. They roughly follow a $A^{-1/3}$ scaling (i.e., the solid line in the figure) but the trend is not smooth as that in $V_{pn}$.

To understand the influence of the large fluctuations in the pairing interaction on $V_{pn}$,
we rewrite Eq. (\ref{ee}) as
\begin{eqnarray}
\nonumber V_{pn} {(Z, N)}&=&\frac{1}{2}\left[S_{n}(Z,N-1)-S_{n}(Z-2,N-1)\right]\\
&&+\frac{1}{4}\left[\Delta_n(Z,N)-\Delta_n(Z-2,N)\right].
\end{eqnarray}
The quantities 
$\delta S_n(Z,N-1)=S_{n}(Z,N-1)-S_{n}(Z-2,N-1)$ and $\delta_n(Z,N)= \Delta_n(Z,N)-\Delta_n(Z,N-2)$
measure the isospin dependences of the one-body separation energy (the mean-field) and pairing 
interaction, respectively. One can easily see that $\delta S_n(Z,N-1)$ (and $\delta S_p(Z-1,N)$) also
reveals the average proton-neutron interaction between the last proton pair and odd neutron as
\begin{eqnarray}\label{vpn-o}
 V_{pn}(Z,N-1)&=&\frac{1}{2}\delta S_n(Z,N-1)\\
\nonumber&=&\frac{1}{2}\left[
B(Z,N-1)+B(Z-2,N-2)\right. \\ 
\nonumber&&- \left.B(Z-2,N-1)-B(Z,N-2) \right].
\end{eqnarray}

\begin{figure}
\centerline{\includegraphics[width=0.65\textwidth]{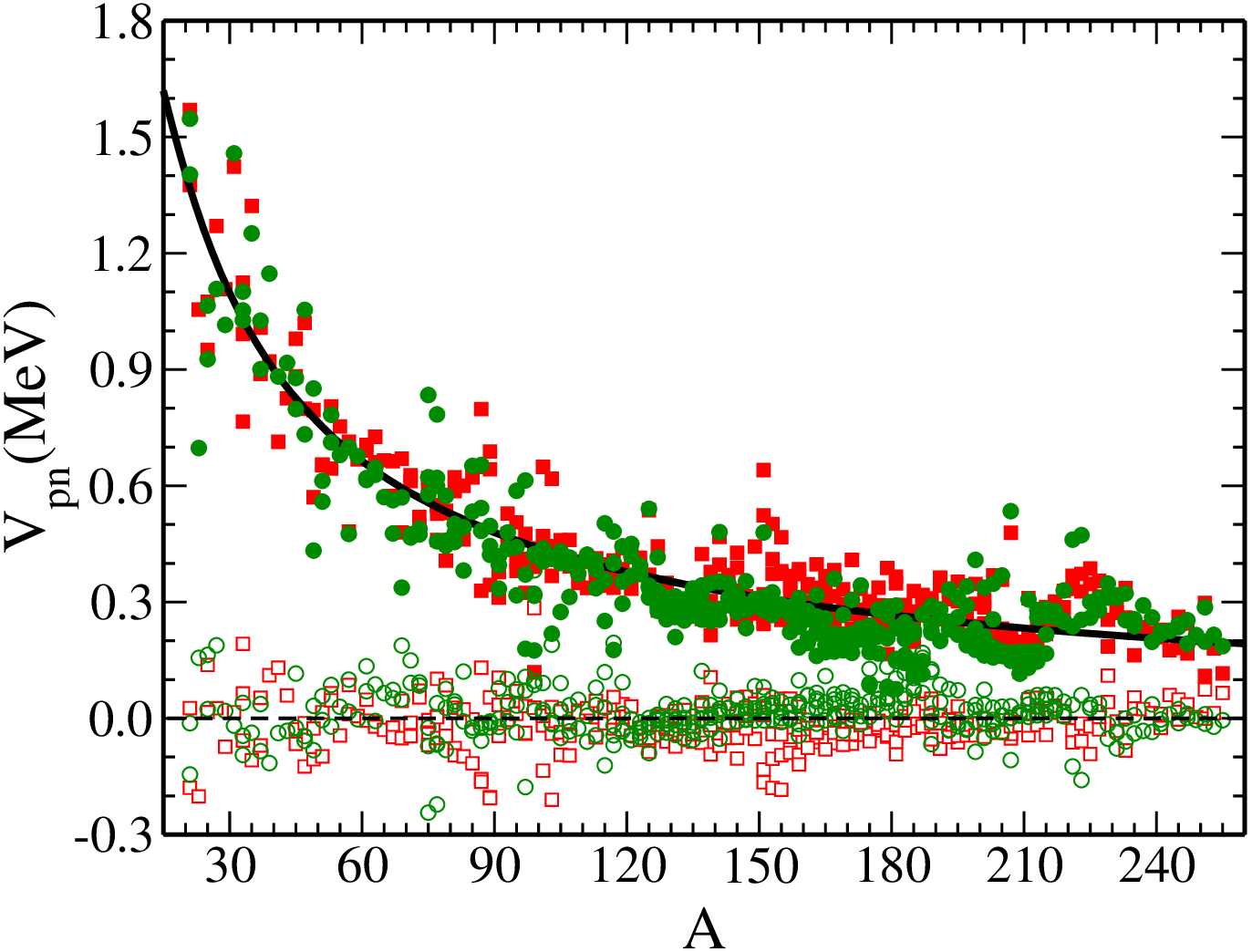}}
\caption{(color online). Empirical proton-neutron interactions in odd-even (Red) and even-odd (Green) nuclei (i.e., $\delta S$)
extracted from experimental nuclear masses~\cite{Audi03,Nei09}. The open symbols correspond to $\delta_p$ 
and $\delta_n$.
The solid line describes
the average behavior of $V_{pn}$ in even-even $N\neq Z$ nuclei.\label{odda}}
\end{figure}

Contributions from the two basic ingredients $\delta S$ and $\delta$ on the empirical proton-neutron 
interaction $V_{pn}$ can be extracted from experimental nuclear masses. The results are plotted in Fig.~\ref{odda}.
It is seen that $V_{pn}$ is dominated by the contribution from $\delta S$. The $\delta_n$ and $\delta_p$ values are 
comparatively small, mostly within $|\delta|\leq 100$~keV. This indicates that the empirical 
proton-neutron interaction
can to a large extent be understood as a mean-field effect. This mechanism is supported by our calculations 
with the DZ and HFB-17 models. It is also consistent with the observation of Ref.~\cite{Sto07}
that HFB calculations on $V_{pn}$ are 
insensitive to the different choices of pairing forces.

if the local fluctuations in the pairing interactions are negligible, it should be
\begin{equation}\label{e-o}
V_{pn}(Z,N)\approx V_{pn}(Z,N-1)\approx V_{pn}(Z-1,N),
\end{equation}
where $Z$ and $N$ are even numbers. This is indeed the case, as can be seen from Fig. \ref{odda}.

\begin{figure}
\centerline{\includegraphics[width=0.65\textwidth]{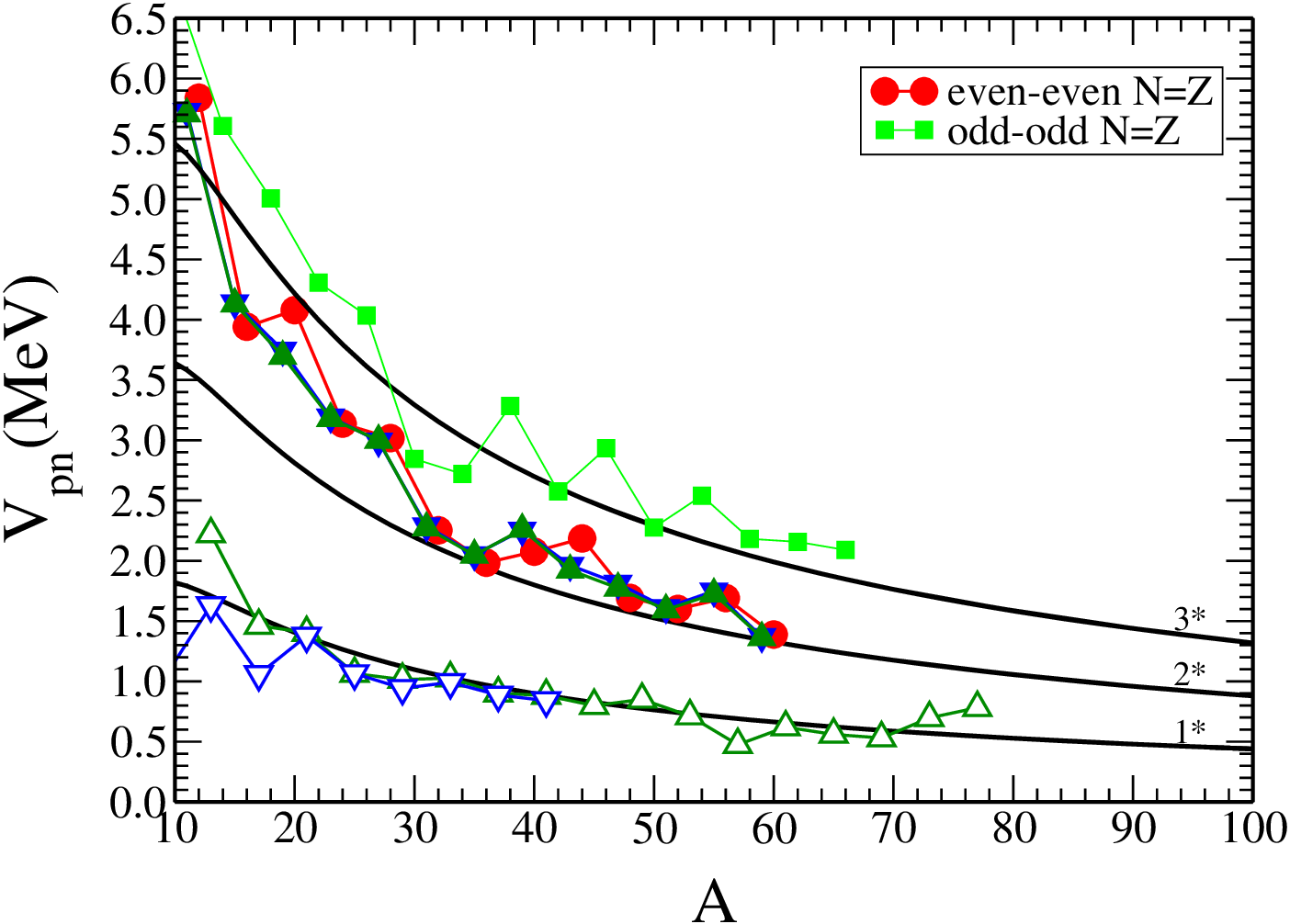}}
\caption{(color online). Experimental $V_{pn}$ values of even-even $N=Z$ nuclei (filled circles) and the adjacent  odd-odd (squares) and odd-$A$ nuclei (triangles). The filled and open triangles correspond to systems with one nucleon subtracted from
and added to the even-even nuclei, respectively. The solid line labeled 1* describes
the average behavior of $V_{pn}$ in even-even $N\neq Z$ nuclei  from Fig. \ref{ee}. 2* and 3* denotes its twice and three time values.\label{nz}}
\end{figure}

In Figs.~\ref{ee} \& \ref{odda} only nuclei with $N\neq Z$ are plotted. 
The extracted proton-neutron 
interactions $V_{pn}$ in $N=Z$ nuclei are noticeably larger than those of the adjacent $N\neq Z$ nuclei, as seen in Fig. \ref{nz}, indicating 
that there is an additional binding in these self-conjugate nuclei. 
This is often described as the Wigner effect \cite{Satu97,Zeldes98,Kir08,Cas10}. Its origin 
has been intensively investigated in the past decade in terms of proton-neutron pairing correlation \cite{Satu97} and spin-isospin symmetry~\cite{Isa95}.
Ref. \cite{Satu97} found that the Wigner effect can not be explained in term of $J=1$ isoscalar neutron-proton pair correlations.

To analyze this feature we consider a system with $n_{\pi}$ protons and $n_{\nu}$ neutrons in a single-$j$ shell.
We assume that the two-body interaction 
obeys a simple form \cite{Talmi93}
\begin{equation}
\hat{V}=a+b\mathbf{t}_1\cdot\mathbf {t}_2+GP_0,
\end{equation}
where $P_0$ denotes the monopole pairing interaction. $G$ is the corresponding (negative) coupling strength. The first two terms, which do not depend on the angular momentum $J$, define the ``averaged" monopole interaction. The isovector and isoscalar channels of the monopole interaction are given by
\begin{equation}
V_{m;T=1}=a+b/4,
\end{equation}
and
\begin{equation}
V_{m;T=0}=a-3b/4.
\end{equation}
 In usual shell-model Hamiltonians the values of $V_{m;T=1}$ are around zero while those of
$V_{m;T=0}$ are strongly attractive (see, e.g., Refs.~\cite{USD,hon09}), indicating that $b$ should have a positive sign \cite{Mek12}.
The $J=0$ two-body matrix element is given as $\langle j^2|V|j^2\rangle_{J=0,T=1}=a+b/4+(2j+1)G$.
The total energy of the system can be written analytically as
\begin{eqnarray}\label{ene}
E=\varepsilon n + \frac{a}{2}n(n-1)+\frac{b}{2}\left[\mathcal{T}(\mathcal{T}+1)-\frac{3n}{4}\right]\\
\nonumber+G\left[\frac{n-v}{4}(4j+8-n-v)-\mathcal{T}(\mathcal{T}+1)+s(s+1)\right],
\end{eqnarray}
where $\varepsilon$ denotes the single-particle energy. The total number of nucleon pairs is $n(n-1)/2$ with $n=n_{\pi}+n_{\nu}$ \cite{Talmi93}. $\mathcal{T}$ is the total isospin of the system. $v$ and $s$ denote the seniority and the reduced isospin. 

For the ground state of an even-even nucleus we have $\mathcal{T}=|n_{\pi}-n_{\nu}|/2$, $s=0$ and $v=0$.
For even-even nuclei with $n_{\pi}\neq n_{\nu}$, we have
\begin{equation}\label{pnn}
V_{pn}=-\frac{4V_{m;T=1}+2(V_{m;T=0}-V_{m;T=1})}{4}=\frac{b}{4}  -a.
\end{equation}
The minus sign in the first term takes into account that the binding energy and $V_{pn}$ are defined as positive in the present work. On the other hand, in the case of $n_{\pi}= n_{\nu}$ (i.e., $N=Z$), we have
\begin{eqnarray}\label{pne}
\nonumber V_{pn}&=&-\frac{4V_{m;T=1}+3(V_{m;T=0}-V_{m;T=1})}{4}-\frac{G}{2}\\
&=&\frac{b}{2}-a -\frac{G}{2}.
\end{eqnarray}

The difference between the $V_{pn}$ in Eqs. (\ref{pnn}) and (\ref{pne}) is $-(V_{m;T=0}-V_{m;T=1})/4-G/2$ or $b/4-G/2$.
The large $V_{pn}$ values would result in a sudden kink in the one-body separation energy $S$ 
when approaching the $N=Z$ line (c.f., Eq.~(\ref{vpn-o})), which can not be reproduced by usual mean-field calculations~\cite{Chas07}. 

\begin{figure}
\centerline{\includegraphics[width=0.45\textwidth]{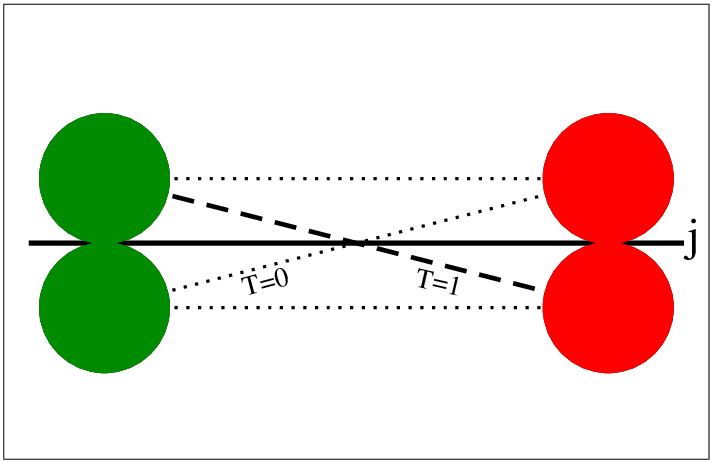}}
\caption{(color online). The $T=1$ (dashed line) and $T=0$ (dotted line) monopole interactions between a neutron pair and a proton pair in a single-$j$ shell. We have a total number of four interacting pairs among which only one has $T=1$. The isovector proton-neutron pairing matrix element also contributes in the $T=1$ channel. \label{sch}}
\end{figure}

A schematic picture is plotted in Fig. \ref{sch} to understand further the coupling of protons and neutrons in $N=Z$ nuclei. For two proton-neutron pairs in a single-$j$ shell, we have three $T=0$ interaction pairs and one $T=1$ interaction pair. The $T=1$ pairing matrix element also contributes to the total binding energy.
For a system with two neutrons and two protons in different shells we have two interaction pairs for both the $T=0$ and 1 
 channels \cite{Hey94}. 
 
 For a $I=j$, $T=1/2$ system with three particles in a single-$j$ shell, we have $v=1$ and $s=1/2$. The $V_{pn}$ for such a nucleus can be expressed in the same form as above. The empirical relation of Eq.~(\ref{e-o}) still holds for these self-conjugate nuclei. In reality we have
\begin{equation}
V_{pn}(Z,Z)\approx V_{pn}(Z,Z-1)\approx V_{pn}(Z-1,Z),
\end{equation} 
where $Z$ takes even values.
 
The empirical interactions between the odd proton and odd neutron in odd-odd nuclei can be extracted from binding energies
in a way similar to those of even-even and odd-$A$ systems. The ground state of odd-odd $N=Z$ nuclei may carry isospin quantum numbers $T=0$ or 1. For the lowest $T=0$ state one may extract the proton-neutron interaction as
\begin{eqnarray}\label{pno}
\nonumber &&V_{pn}(Z-1,Z-1) \\
\nonumber &&= B(Z-1,Z-1)+B(Z-2,Z-2)\\
\nonumber&& -B(Z-1,Z-2)-B(Z-2,Z-1)\\
&&=\frac{3b}{4}-a.
\end{eqnarray}
The results are also plotted in Fig. \ref{nz}. 

Fig. \ref{nz} indicates that the $V_{pn}$ in even-even $N=Z$ nuclei and the adjacent odd-$A$ nuclei with one less nucleon are roughly twice as large as those in neighboring $N\neq Z$ nuclei, while the $V_{pn}$ in odd-odd $N=Z$ nuclei are three times as large as the average values in $N\neq Z$ nuclei. This may be understood from Eqs. (\ref{pnn}), (\ref{pne}) and (\ref{pno}) by assuming that $a\sim 0$.  The figure suggests that in reality $b$ should be positive. In medium mass and heavy nuclei, it should also be much larger than the pairing strength $G$. This is consistent with the results of empirical shell model calculations \cite{USD,hon09}. The $V_{pn}$ in very light even-even $N=Z$ nuclei are much larger than the average behavior . This may be due to the additional binding gained from the enhanced pairing energy which scales like $A^{-1/3}$.
In the spin-isospin SU(4) symmetry limit, the $V_{pn}$ of $N=Z$ nuclei are four times larger than those for $N\neq Z$ \cite{Isa95}. There was also no difference between $V_{pn}$ in even-even and odd-odd $N=Z$ nuclei \cite{Isa95}.

 We made no attempt to fit the strengths of the monopole interactions for single-$j$ systems. In practice, for $0d_{5/2}$ nuclei with $A\sim20$, we have $V_{m;T=0}\sim-5.5$ MeV \cite{USD} and $G\sim-0.8$ MeV. 
  For $0f_{7/2}$ nuclei with $A\sim50$, we have $V_{m;T=0}\sim-1.6$ MeV \cite{hon09}. These values reproduce reasonably the differences between $V_{pn}$ in $N=Z$ and $N\neq Z$ nuclei. One may expect that the ``residual" proton-neutron interaction beyond the isoscalar monopole interaction also influence the $V_{pn}$. This residual interaction, which is more complicated since it breaks the seniority symmetry \cite{Qi11}, is not taken into account in the present work.

On the first glance one may say that the isovector proton-neutron pairing also has significant influence on the $V_{pn}$ for even-even $N=Z$ nuclei.
But in real cases the pairing interaction (and other matrix elements)
is strongly modified by the monopole interaction. To evaluate the residual effect of the pairing interaction we rewrite Eq. (\ref{ene}) as
\begin{eqnarray}
E&=&\varepsilon n + \frac{2a-G}{4}n(n-1)\\
\nonumber&&+\frac{b-2G}{2}\left[\mathcal{T}(\mathcal{T}+1)-\frac{3n}{4}\right]\\
\nonumber&&+(j+1)G(n-v)+G\left[\frac{v^2}{4}-v+s(s+1)\right],
\end{eqnarray}
from which it can be seen that it is the term $(j+1)Gv$ that may result in an odd-even staggering in nuclear binding energies. This suggested that the residual pairing term in macroscopic mass formulas may be written as
\begin{equation}\label{pt}
E_p\propto 2-v,
\end{equation}
where $v=1$ for odd-$A$ nuclei and $v=2$ for the $\mathcal{T}=|N-Z|$/2 ground state of odd-odd nuclei. There should be no additional gain in pairing energy when crossing the $N=Z$ line. This is consistent with the results in Fig. \ref{pair} where no noticeable difference is seen between the pairing energies of $N=Z$ and other nuclei.

The DZ mass model is constructed starting from a shell-model monopole Hamiltonian as
\begin{equation}\label{dz}
H_m= H_M+H_s + H_d,
\end{equation}
where $H_M$ is the macroscopic part including the symmetry energy (proportional to $\mathcal{T}(\mathcal{T}+1)$) and the pairing energy. 
The microscopic spherical term $H_s$ and the deformed term $H_d$ take into account the residual three-body and four-body correlations between
valence nucleons in the open shell. 
The expectation value of the Hamiltonian $H_s$ is calculated by assuming the normal filling scheme of nucleons.
The deformed Hamiltonian $H_d$ takes into account the effect of the promotion of valence nucleons to
the next shell~\cite{Duf95}. 
The DZ model also contains a phenomenological $\mathcal{T}(\mathcal{T}-1/2)$ term (referred to as the ``Wigner" term in Ref. \cite{dz}) and a $\mathcal{T}/A$ correction to the pairing energy. These terms have limited influence on $V_{pn}$ and the binding energies.
One can obtain a slightly better agreement with experiments by refitting the DZ model parameters to the up-to-date mass table \cite{Audi03} without these two terms and with the pairing term as Eq. (\ref{pt}). The deviation from experiment is $\sigma=0.537$ MeV. In this simplified model the only term that explicitly depends on isospin $\mathcal {T}$ is the symmetry energy which induces the kink between $V_{pn}$ for $N=Z$ and $N\neq Z$ nuclei as seen in Fig. \ref{nz}.

We calculated the contributions from different terms of the DZ Hamiltonian on $V_{pn}$. It is thus found that $H_M$ reproduce the bulk properties of the $V_{pn}$ while the many-body term $H_s$ may lead to fluctuations around their mean values.

\begin{figure}
\vspace{1.5cm}
\centerline{\includegraphics[width=0.68\textwidth]{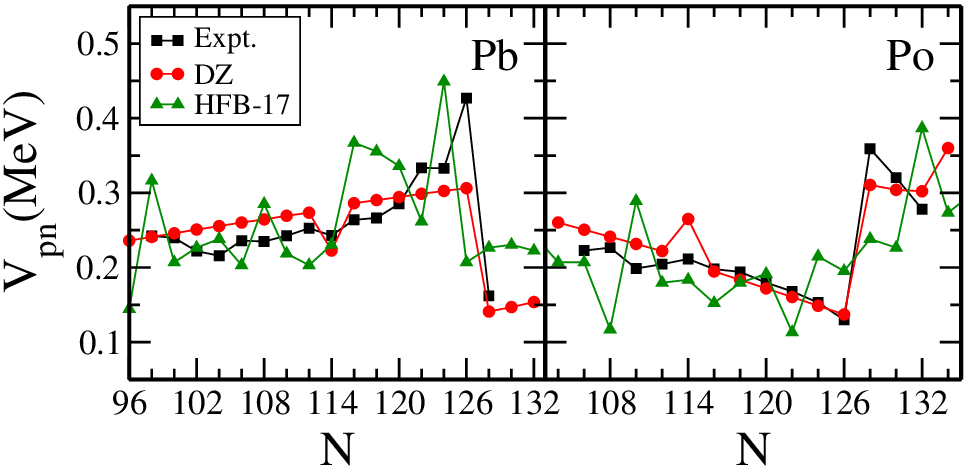}}
\caption{(color online). Experimental and calculated proton-neutron interactions $V_{pn}$ in even-even Pb and Po isotopes. 
Experimental data are taken from Refs.~\cite{Audi03,Chen09}.\label{n126}}
\end{figure}

Recently the $V_{pn}$ in nuclei around $N=126$ and $Z=82$ shell closures have been intensively investigated~\cite{Cas05,Nei09,Chen09}.  Experimental and calculated $V_{pn}$ values in Pb and Po isotopes are plotted in Fig.~\ref{n126}. The $V_{pn}$ values in nuclei below and above $Z=82$ follow similar evolution patterns to those of Pb and Po isotopes, respectively. As can be seen from the figure,
the evolution of $V_{pn}$ shows a clear bifurcation pattern~\cite{Cas05}. That is nuclei below and above $Z=82$ evolve in two distinct ways as a function of neutron number $N$. 
Energy density functional calculations with the SkP Skyrme force cannot describe this behavior~\cite{Sto07}. A better agreement was obtained for Pb isotopes in Ref. \cite{Ben11} by taking into account the influence of beyond-mean-field correlations. Calculations with the DZ and HFB-17 models are also plotted in Fig. \ref{n126} for comparison. The DZ model shows a better agreement with experiments than other calculations.

\begin{figure}
\vspace{1cm}
\centerline{\includegraphics[width=0.68\textwidth]{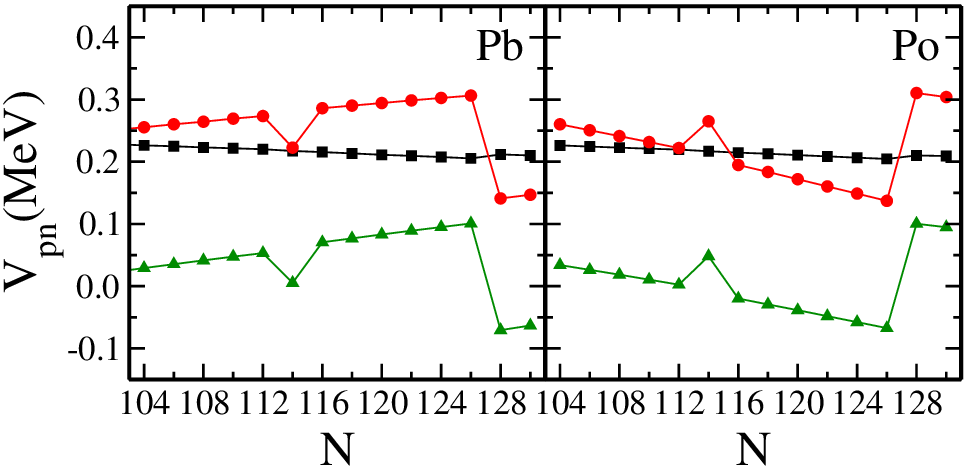}}
\caption{(color online). DZ model calculations (circle) on $V_{pn}$ values of even-even Pb and Po isotopes and the 
contributions from $H_s$ (triangle) and $H_M$ (square) of the monopole Hamiltonian.\label{dz126}}
\end{figure}

Calculations with the DZ model show that the $V_{pn}$ values in this mass region are dominated by contributions from the monopole terms in $H_M$ and $H_s$, as can be seen from Fig. \ref{dz126}.

\begin{figure}
\vspace{1cm}
\centerline{\includegraphics[width=0.65\textwidth]{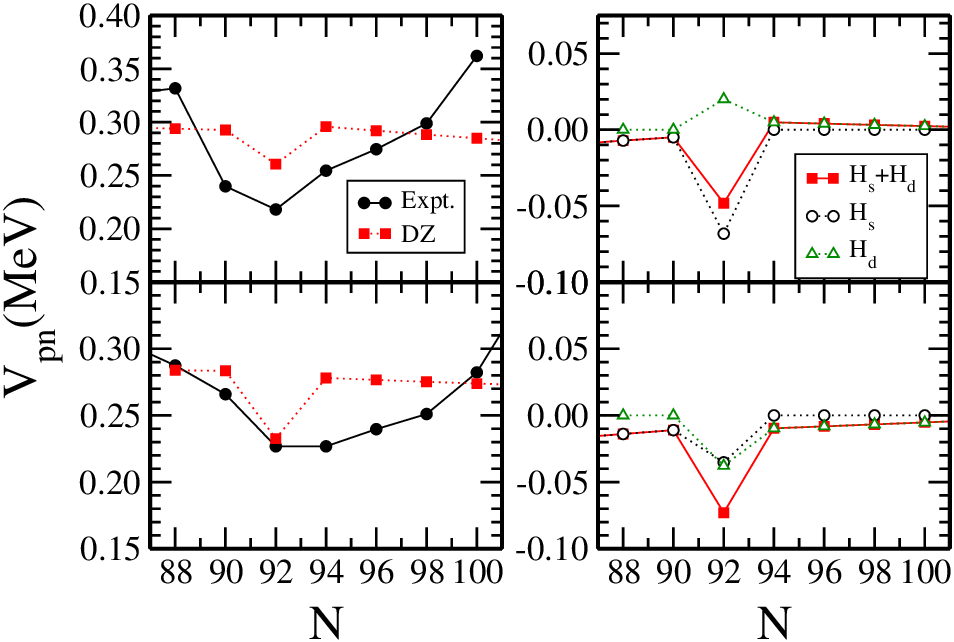}}
\caption{(color online). Left: experimental and calculated proton-neutron interactions $V_{pn}$ in even-even Er 
(upper) and Yb (lower) isotopes. Right: Contributions from $H_s$ and $H_d$ on $V_{pn}$ values.\label{n92}}
\end{figure}

A sudden drop at $N=92$ was noted in the $V_{pn}$ values of Er isotopes, which was explained in terms of transitions from spherical to deformed shapes in Ref.~\cite{Sto07} . 
This is supported by calculations with the HFB-17 and DZ mass models. The results of DZ model calculations are plotted in the left panels of Fig.~\ref{n92}, in comparison with experimental data.
The evolutions of $V_{pn}$ values in Yb and Hf isotopes show similar patterns (in the latter case the location of the drop moves to $N=94$).
In the right panels of the figure we plotted the contributions of the spherical and deformed monopole terms $H_s$ and $H_d$. It can be easily seen that the sudden drop is a result of the competing effect of spherical and deformed monopole terms $H_s$ and $H_d$.

In summary, a systematic analysis on the empirical proton-neutron interaction $V_{pn}$ is done from a simple perspective by describing it as the interplay between the mean field and the pairing interaction.
The results are also compared with those derived from existing nuclear energy density functional and monopole shell-model calculations. 
It is found that the bulk properties of the $V_{pn}$ are dominated by the contribution from mean field which can be estimated empirically from the one-nucleon separation energies. The pairing energy plays a relatively minor role. For the same reason the $V_{pn}$ for odd-$A$ nuclei are close to those of the neighboring even-even nuclei.

The $V_{pn}$ for $N=Z$ nuclei and the odd-$A$ nuclei with one nucleon subtracted from the even-even $N=Z$ ones are much larger than others. We analyzed this feature within a simple seniority model by including the isovector and isoscalar monopole interactions and the monopole pairing interaction. It is thus found that in these nuclei there is
a strong additional contribution from the isoscalar monopole interaction and isovector proton-neutron pairing interaction. As a result, the $V_{pn}$ for even-even and odd-odd $N=Z$ are roughly two and three times stronger than those for $N\neq Z$ nuclei.  In the DZ model the cusp is induced by the $\mathcal{T}$ dependent term in the symmetry energy which is proportional to $\mathcal{T}(\mathcal{T}+1)$.

We also analyzed the local fluctuations in $V_{pn}$ around their mean values. In the Duflo-Zuker mass model, these fluctuations are understood as the interplay of three-body- and four-body-like monopole interactions between valence nucleons in the open shell. As examples, we applied the DZ model to investigate the $V_{pn}$ evolution pattern in nuclei around $N=126$ and $N=92$, where intensive attempts were done within the Hartree-Fock approach.

The author thanks R. Liotta and R. Wyss for stimulating discussions and R. Liotta for his reading of the manuscript.
This work has been supported by the Swedish Research Council (VR) under grant No. 621-2010-4723.


\begin{thebibliography}{50}
\bibitem{bm}A. Bohr and B. R. Mottelson, {\it Nuclear Structure} (World
Scientific, Singapore, 1998).
\bibitem{Ber09}G. F. Bertsch, C. A. Bertulani, W. Nazarewicz, N. Schunck, and M. V. Stoitsov, 
Phys. Rev. C {\bf79} (2009) 034306, and references therein.
\bibitem{Gor09}S. Goriely, N. Chamel, and J. M. Pearson, Phys. Rev. Lett. {\bf102} (2009) 152503; http://www.astro.ulb.ac.be/pmwiki/Brusslib/Hfb17
\bibitem{Ben03} M. Bender, P.-H. Heenen and P.-G. Reinhard, 
Rev. Mod. Phys. {\bf75} (2003) 121.
\bibitem{Lunney} D. Lunney, J. M. Pearson and C. Thibault, Rev. Mod. Phys. {\bf 75} (2003) 1021.
\bibitem{Fed77}P. Federman and S. Pittel, Phys. Lett. {\bf B69} (1977) 385.
\bibitem{Cas06}R. B. Cakirli and R.F. Casten, Phys. Rev. Lett. \textbf{96} (2006) 132501.
\bibitem{Otsuka01}T. Otsuka, R. Fujimoto, Y. Utsuno, B. A. Brown, M. Honma, and T. Mizusaki, 
Phys. Rev. Lett. {\bf87} (2001) 082502.
\bibitem{Sor08}O. Sorlin, M.G. Porquet, Prog. Part. Nucl. Phys. 61 (2008) 602.
\bibitem{Zhang89} J.-Y. Zhang, R. F. Casten and D. S. Brenner, 
Phys. Lett. \textbf{B227} (1989) 1.
\bibitem{Cas05}R. B. Cakirli, D. S. Brenner, R.F. Casten and E.A. Millman,
Phys. Rev. Lett. \textbf{94} (2005) 092501.
\bibitem{Cas06a}Y. Oktem, R. B. Cakirli, R. F. Casten, R. J. Casperson, and D. S. Brenner, 
Phys.~Rev.~\textbf{C 74} (2006) 027304.
\bibitem{Cas06b}D. S. Brenner, R. B. Cakirli, and R. F. Casten, Phys. Rev. \textbf{C 73} (2006) 034315.
\bibitem{Sto07}M. Stoitsov, R. B. Cakirli, R. F. Casten, W. Nazarewicz, and W. Satu{\l}a, 
Phys. Rev. Lett. {\bf98} (2007) 132502.
\bibitem{Fu10}G. J. Fu, Hui Jiang, Y. M. Zhao, and A. Arima,
Phys. Rev. C 82 (2010) 014307.
\bibitem{Jiang12}H. Jiang, G. J. Fu, Y. M. Zhao, and A. Arima, Phys. Rev. C 85 (2012) 024301.
\bibitem{Macc00}A.O. Macchiavelli {\textit et al.}, Phys. Rev. C {\bf61} (2000) 041303(R).
\bibitem{Good01}A.L. Goodman, Phys. Rev. C {\bf63} (2001) 044325.
\bibitem{Satu01}W. Satu{\l}a and R. Wyss, Phys. Rev. Lett. {\bf86} (2001) 4488; 
Phys. Rev. Lett. {\bf87} (2001) 052504.
\bibitem{Chas07}R.R. Chasman, Phys. Rev. Lett. {\bf99} (2007) 082501.
\bibitem{Qi11}C. Qi, J. Blomqvist, T. B\"ack, B. Cederwall, A. Johnson, R. J. Liotta, and R. Wyss,
Phys. Rev. C 84 (2011) 021301.
\bibitem{Ben11}M. Bender and P.-H. Heenen, Phys. Rev. C 83 (2011) 064319.
\bibitem{Audi03} G. Audi, A. H. Wapstra and C. Thibault, 
Nucl. Phys. \textbf{A729} (2003) 337; http://amdc.in2p3.fr/masstables/filel.html
\bibitem{Satu97} W.~Satu{\l}a, D.J. Dean, J. Gary, S. Mizutori, W. Nazarewicz, Phys.~Lett.~{\bf B407} (1997) 103.
\bibitem{Duf95}J. Duflo and A.P. Zuker, Phys. Rev. C {\bf52} (1995) R23. 
\bibitem{dz}http://amdc.in2p3.fr/web/dz.html
\bibitem{Satu98}W. Satu{\l}a, J. Dobaczewski, and W. Nazarewicz, Phys. Rev. Lett. {\bf81} (1998) 3599.
\bibitem{Nei09}D. Neidherr {\it et al.}, Phys. Rev. Lett. {\bf102} (2009) 112501.
\bibitem{Zeldes98}N. Zeldes, Phys. Lett. B 429 (1998) 20. 
\bibitem{Kir08}M.W. Kirson, Phys. Lett. B 661 (2008) 246.
\bibitem{Cas10}R. B. Cakirli, D. S. Brenner, R.F. Casten, Phys. Rev. C 83 (2010) 061304(R).
 \bibitem{Isa95}P. Van Isacker, D.D. Warner, and D.S. Brenner, Phys. Rev. Lett. {\bf74} (1995) 4607.
\bibitem{Talmi93}I. Talmi, \textit{Simple Models of Complex Nuclei} (Harwood Academic Publishers, Chur, Switzerland, 1993).

\bibitem{USD}B. A. Brown
and W. A. Richter, Phys. Rev. C {\bf74} (2006) 034315.
\bibitem{hon09}M. Honma, T. Otsuka, B. A. Brown, and T. Mizusaki
Phys. Rev. C 69 (2004) 034335.

\bibitem{Mek12}A.Z. Mekjian and L. Zamick, Phys. Rev. C {\bf85} (2012) 057303.
\bibitem{Hey94}K. Heyde, C. De Coster and J. Schietse, Phys. Rev. C {\bf49} (1994) 2499.

\bibitem{Chen09}L. Chen {\it et al.}, Phys. Rev. Lett. {\bf102} (2009) 122503.
\end{thebibliography}
\end{document}